\documentclass[aps,amssymb,amsmath,amsfonts,twocolumn,pra]{revtex4}
\usepackage[latin1]{inputenc}
\usepackage{amsthm} 
\usepackage{amsmath}
\usepackage{graphicx} 
\usepackage{subfigure} 
\usepackage{amsfonts}
\usepackage{epstopdf}
\usepackage{hyperref} 
\usepackage{xcolor}
\usepackage{verbatim} 
\usepackage{bm,bbm}
\usepackage{bbold}

\newcommand{\bra}[1]{\langle #1|}
\newcommand{\floor}[1]{\lfloor #1 \rfloor}
\newcommand{\ket}[1]{|#1\rangle}

\newcommand{\ketbra}[2]{| #1 \rangle \langle #2 |}

\newcommand{\proj}[1]{\vert #1\rangle\!\langle#1 \vert}

\newcommand\Tr{{\mathrm{Tr}}}
\newcommand\iu{{\mathrm{i}}}
\newcommand\e{{\mathrm{e}}}

\begin{document}
\title{Absolutely Maximally Entangled states,\\
  combinatorial designs and multi-unitary matrices}

\author{Dardo Goyeneche}
\affiliation{National Quantum Information Center of Gda\'{n}sk,  81-824 Sopot,  Poland}
\affiliation{Faculty of Applied Physics and Mathematics, Technical University of Gda\'{n}sk, 80-233 Gda\'{n}sk, Poland}
\author{Daniel Alsina}
\address{Dept. Estructura i Constituents de la Mat\`eria, Universitat de Barcelona, Spain.}
\author{Jos\'e~I.~Latorre}
\address{Dept. Estructura i Constituents de la Mat\`eria, Universitat de Barcelona, Spain.}
\address{Center for Theoretical Physics, MIT, USA}
\author{Arnau Riera}
\address{ICFO-Institut de Ciencies Fotoniques, Castelldefels (Barcelona), Spain}
\author{Karol {\.Z}yczkowski}
\address{Institute of Physics, Jagiellonian University, Krak\'ow, Poland}
\address{Center for Theoretical Physics, Polish Academy of Sciences, Warsaw, Poland}

\date{June 29, 2015}

\begin{abstract}
Absolutely Maximally Entangled (AME) states are those multipartite quantum states that carry  absolute maximum entanglement in all possible partitions.
 AME states are known to play a relevant role in multipartite teleportation, 
in quantum secret sharing and they provide the 
basis novel tensor networks related to holography. 
We present alternative constructions of 
AME states and show their link with combinatorial designs. We also
analyze a key property of AME, namely their relation to tensors that
can be understood as unitary transformations in every of its bi-partitions.
We call this property multi-unitarity.
\end{abstract}

\maketitle
 
\section{Introduction}\label{S1}
A complete characterization, classification and quantification of entanglement 
for quantum states remains an unfinished long-term goal in Quantum Information theory. Nevertheless, a large number of relevant results related to entanglement are well-known by now. It is, for instance, known that generic condensed-matter physical systems characterized by homogeneous nearest neighbor interactions carry an amount of entropy bounded by an area law. This fact opens the possibility of applying tensor network techniques to describe the ground state of relevant physical systems, including such phenomena as quantum phase transitions.

However, there are many situations where entanglement entropy is maximal or nearly maximal, that is, it scales as the volume of the quantum system. This is the  case of generic time evolution of local Hamiltonians, of evolution of a quantum computer addressing a quantum Merlin--Arthur (QMA) 
problem, and of random states. The latter example is useful for our discussion, since it is known that the average entanglement of a random pure state $|{\psi}\rangle$ of $N$ qubits is close to maximal. Indeed, making use of the von Neumann entropy
\begin{equation}
S(\rho)=-\Tr (\rho \log \rho)\, ,
\end{equation}
 it is possible to show \cite{P93, BZ06} that the average entropy of the reduced state $\sigma=\Tr_{N/2}|\psi\rangle\langle\psi|$ to $N/2$ qubits reads:
\begin{equation}
  S(\sigma)=\frac{N}{2}\log2-c+({\rm subleading\ terms}),
\end{equation}  
where the constant $c$ depends on the choice of the ensemble 
used to generate random states \cite{sen96, ZS01}. For convenience we will consider natural logarithm along the work.
In the case the average is defined with respect to
the unitarily invariant Haar measure on the space of unit
vectors of size $2^N$
the constant reads $c=1/2$ \cite{P93}.
Although random states are {\sl almost} maximally entangled, 
their entropy differs from the maximal value by a negative constant.
One  may then investigate when truly maximally entangled states appear, what properties they have and what they are useful for. A relevant motivation for this study is
the role that absolute maximal entanglement may play in the context of holography, as we shall discuss later on.


The aim of the present paper is to extend the relation of absolutely maximally entangled (AME) states to a different branch of mathematics, so-called combinatorial designs.
On one hand, new families of maximally entangled multipartite states are presented.
On the other hand, we demonstrate a direct link between multipartite entanglement
and combinatorial structures, including mutually orthogonal Latin squares and Latin cubes, 
and symmetric sudoku designs. 
Furthermore, we introduce the concept of \emph{multiunitary matrices},
which exists in power prime dimensions, show their usage in constructing AME states 
and present a small catalog of such matrices in small dimensions.

This work is organized as follows. In Section \ref{S2} we present some important aspects of entanglement theory, AME states and its link with holography. In Section \ref{S3} we review the current state of the art of AME states for multipartite systems. In Section \ref{S4} we consider the particular case of AME states having minimal support and its relation to classical codes and orthogonal arrays of index unity. A link between AME states, mutually orthogonal Latin squares and Latin cubes and hypercubes is explored in Section \ref{S5}. In Section \ref{S6} we introduce the concept of multiunitary matrices and demonstrate that they are one-to-one connected with AME states. Additionally, we provide some constructions of multi-unitary matrices and its associated AME states. In Section \ref{S7} we resume and discuss the most important results obtained in our work and conclude. In Appendix \ref{appendixA} we discuss the non-existence of AME states of four qubits and present the most entangled known states. In Appendix \ref{appendixB} we detailedly explain the simplest case of 2-unitary matrices of order $D=d^2$, which are associated to AME states of 4 qu\emph{d}its. some examples of multiunitary matrices are presented. In Appendix \ref{appendixC} we establish a relationship between AME states and a special class of Sudoku. Finally, in Appendix \ref{appendixD} we present a mini-catalog of multi-unitary matrices existing in low dimensions.

\section{AME states: definition and basic properties}\label{S2}

\subsection{Definition}
A lot of attention was recently paid to identify entangled states of $N$--party systems,
such that tracing out arbitrary $N-k$  subsystems, the remaining $k$ subsystems are maximally mixed \cite{GBP98,HS00,Brown05,Facchi2008,FFMPP10b,AC13}.
Such states are often called $k$--uniform \cite{Sc04,GZ14}
and by construction the integer number $k$ cannot exceed $N/2$.
In this paper, we shall focus on the extremal case, $k=\floor{N/2}$ (we put the floor function $\floor{}$ to include cases of N even and odd) ,
and analyze properties of states called
{\sl absolutely maximally entangled} (AME) -- see \cite{HCLRL12,He13}. Also, such states were previously known as perfect maximally multipartite entangled states \cite{Facchi2009}.

The definition of AME states corresponds to those quantum states that carry maximum entropy in {\sl all} their bi-partitions.
 It is a remarkable fact that the existence of  
such states is not at all trivial and deepens into several branches of mathematics. Let us be more precise and  define an AME($N$,$d$) state
$|\psi\rangle \in {\cal H}$, made with $N$ qudits of local dimension $d$, 
 ${\cal H}=(C^d)^{\otimes N}$ as a state such that all its reduced density matrices in any subspace
${\cal A}=(C^d)^{\otimes \frac{N}{2}}$, ${\cal H}={\cal A} \otimes \bar{\cal A}$,
carry maximal entropy
\begin{equation}
  S(\rho_{\cal A})=\frac{N}{2} \log d \qquad \forall {\cal A}  \ .
\end{equation}
This is tantamount to asking the reduced density matrices to $k$ qudits to be
proportional to the identity
\begin{equation}
\rho_k=\frac{1}{d^k} I_{d^k} \qquad \forall k\leq \frac{N}{2}\ .
\end{equation}
Let us note the fact that a $k$--uniform state is also $k^{\prime}$--uniform for any $0<k^{\prime}<k$.


There is an obstruction for a state to reach maximal entanglement in all partitions
due to the concept of monogamy of entanglement \cite{CKW00,T03}. Every local degree of freedom
that tries to get maximally entangled with another one is, then, forced to disentangle
from any third party. Therefore, entanglement can be seen
as a resource to be shared with other parties. If two local degrees of freedom get largely entangled among themselves, 
then they are less able to be entangled with the rest of the system. But this rule is not always fulfilled. There are
cases  where the values of the local dimensions  $d$ and the total number of
qudits $N$ are such that AME states exist. 
For a given $N$, there is always a large enough $d$ 
for which there exists an AME state \cite{HCLRL12}. 
However, the lowest value of $d$ such that an AME state exists is not known in general.

Let us mention that AME states are 
 useful and necessary to accomplish 
certain classes of multipartite protocols. In particular, in Ref. \cite{HCLRL12},
it was shown that AME states are needed to implement two different categories of 
protocols. First, they are needed to achieve perfect multipartite
teleportation. Second, they provide the
resource needed for quantum secret sharing. 
These connections hint at further
relations between AME states and different branches of Mathematics.
For instance, AME states are related to Reed-Solomon codes \cite{RS60}; 
Also, AME states (and $k$--uniform states in general) are deeply linked to error correction codes \cite{Sc04}.

There is yet another surprising connection between AME states and holography \cite{LS15,PYHP15}. It can be seen that AME states provide the basis for a tensor network structure that distributes entanglement in a most efficient and isotropic way. This tensor network can be proven to deliver holographic codes, that may be useful as quantum memories and as microscopic models for quantum gravity. 
A key property for these new developments is related to the
properties of multi-unitarity that will be explored in Sect. \ref{sec:multi-unitarity}.

\subsection{Local Unitary equivalence}

Entanglement is invariant under choices of local basis. It is then
natural to introduce the concept of Local Unitary (LU) equivalence among AME states. 
Two quantum states $\ket{\Phi}$ and $\ket{\Psi}$ are called LU-equivalent
if there exist $N$ local unitary matrices, $U_1,\ldots U_N$ such that
\begin{equation}
\ket{\Phi}= U_1 \otimes U_2 \otimes \ldots \otimes U_N \ket{\Psi} \, .
\end{equation}
If $\ket{\Psi}$ is a maximally entangled state any other state 
LU-equivalent to it is also maximally entangled. 
In this respect, we will define AME($N,d$) as the set of all AME states in the
Hilbert space ${\mathcal H}(N,d)$  and denote their elements by a Greek letter,
e.~g.\ $\ket{\Omega_{4,3}}\in \textrm{AME}(4,3)$ is an AME of four qutrits.

The LU transformations introduce equivalence classes of states. 
A question arises naturally about which state should be chosen as the representative of the class,
that will be denoted as \emph{canonical form} of an AME state. 
It is possible to 
argue in two different directions. On one hand, we may consider that a
natural representative may carry all the elements of the computational
basis. It would then be necessary to establish theorems and a criterion
to fix the coefficients. On the other hand, an alternative possibility is to choose the element of the class with a minimal support on the computational basis. Results in both directions are presented in Sec.~\ref{S4}.

It is not known in general how many different LU classes there are in the set AME($N,d$) for every $N$ and $d$. This question can be tackled by the construction of LU-invariants. 
A few examples are at hand for few qubits. For three qubits, it is known how to obtain
a canonical form of any state using LU and that all
states are classified by 5 invariants \cite{Acin00}, only one of them
is genuinely multipartite, the tangle. For four qubits, there are ways
to construct a canonical form and to find the hyperdeterminant, as well.
Yet, it is unknown how to proceed to larger local dimensions and number
of parties. It is arguable that the subset of AME states is
characterized by several LU-invariants, probably related to distinct
physical tasks. In such a case, there would be different AME states not
related to each other by LU.

\subsection{AME and holography}

Quantum holography amounts to the fact that the information content of a quantum system is that of its boundary. It follows that the information present in the
system is far less than the maximum allowed. Degrees of freedom in the bulk will not carry maximal correlations, neither the
von Neumann entropy of any sub-part of the system will scale
as its volume.

To gain insight in quantum holography, it is natural to investigate the bulk/boundary correspondence of the operator content of the theory. On the other
hand, Quantum Information brings a new point of view on this issue, since
it focuses on the properties of states rather than on the dynamics
that generates it. In this novel context, we may ask what is
the structure of quantum states that display holographic properties.
That is, we aim at finding which
is the detailed entanglement scaffolding that guarantees that
information flows from
the boundary to the bulk of a system in a perfect way.

A concept separate from holography turns out to be very useful to address the
analysis of holography from this new Quantum Information perspective, that is Tensor Networks of the kind of
matrix product states (MPS), projected entangled pair states (PEPS) 
and
 multi-scale entanglement renormalization ansatz (MERA) 
 Indeed, Tensor Networks
provide a frame to analyze how correlations get
 distributed in quantum states, and thus to understand holography
at the level of quantum states. Each connection among ancillary indices
quantifies the amount of entanglement which links parts of the system.
Holography must necessarily relay on some very peculiar entanglement
structure.

A first attempt to understand the basic property behind holography
of quantum states was presented in
 \cite{LS15}. There, it was proposed to create a quantum state on a
triangular lattice based on
a tensor network that uses as ancillary states an absolutely maximally entangled
state. To be precise, the state $\ket{\Omega} \in AME(4,3)$ (see Eq.(\ref{ame43}) was defined on tetrahedrons, in such a
way that the vertices in its basis connect the tensor network and the tip
of the tetrahedron corresponds to a physical index.

In a subsequent work \cite{PYHP15}, another construction was based on
the 5-qubit $\ket{\Upsilon_{5,2}} \in AME(5,2)$ state \ref{ame52}. Again, the fact that the internal construction
of the state is based on isometries is at the heart of the holographic property.

There are two obvious observations on the surprising relation between
AME states and holography. The first is related to the natural link between
AME states and error correction codes. It is arguable, then, that
the essence of holography is error correction, which limits the amount
of information in the system. The second is that the very property
responsible for holography is multi-unitarity, which is analyzed in depth in section \ref{S6}. It is further arguable
that multi-unitarity is the building block of symmetries, since the
sense of direction is lost and can be defined at will. Those ideas deserve
a much deeper analysis.

\subsection{ Related definitions}
\subsubsection{Maximally Entangled sets}
In the context of convertibility of states via
Local Operations and Classical Communication (LOCC), 
multipartite entanglement is significantly different from the bipartite one.
While in the bipartite case, 
there is a single maximally entangled state 
(up to local unitaries)
that can be transformed into any other state by LOCC (and cannot be obtained from any other),
in the multipartite scenario this is no longer true.
In Ref.~\cite{VSK13}, the notion of the Maximally Entangled (ME) set of $N$-partite states is
introduced as the set of states from which any state outside of it can be obtained via LOCC from one of the states within the set and no state in the set can be obtained from any other state via LOCC. 
Note that this notion of maximal entanglement is strictly weaker than the AME, in the sense that
most of (or all) states in the ME set will not be an AME state,
 but any AME state will be in its corresponding ME set. 
In Ref.~\cite{VSK13}, the ME set is characterized for the case of three and four qubits. 
It is interesting to point out that, unlike the 3-qubit case, deterministic LOCC transformations are almost never possible among fully entangled 4-partite states. 
As a consequence of this,
while the ME set is of measure zero for 3-qubit states, almost all states are in the 4-qubit ME set.
This suggests the following picture; given a fixed local dimension and for an increasing number of parties, 
the AME states become more an more rare at the same time that more and more states need to be included
in the ME set. In other words, while maximally entangled states defined from an operational point of view become typical when the number of parties increases, AME states are exotic.

The issue of inter-convertibility between quantum states has been addressed under a larger  set of operations than the LOCC: the Stochastic Local Operations and Classical Communication
(SLOCC). The SLOCC identify states that can be
interconverted by LOCC in a non-deterministic way, but with a non-zero probability of success.
In this respect, a systematic classification of multipartite entanglement in terms of equivalence classes of states under SLOCC is presented in Ref.~\cite{Gour2013}. In particular, it is shown that the SLOCC equivalence class of multipartite states is characterized by ratios of homogeneous polynomials that are invariant under local action of the special linear group.
This work generalizes for an arbitrary number of partitions and finite local dimension the complete classification made for 4-qubits in Ref.~\cite{Verstraete02}.

\subsubsection{Maximally Multipartite Entangled states}

In Ref.~\cite{Facchi2008}, \emph{Maximally Multipartite Entangled} (MME) states are
introduced as those states that maximize the average entanglement (measured in terms of purity) 
where the average is taken over all the balanced bipartitions i.~e.~ $|\mathcal{A}|=\floor{N/2}$.
More specifically, the MME states are defined as the minimizers of the \emph{potential of multipartite entanglement}
\begin{equation}
\pi_{ME}=\binom{N}{\floor{N/2}} \sum_{|\mathcal{A}|=\floor{N/2}}\pi_\mathcal{A}\, ,
\end{equation}
where $\pi_\mathcal{A}=\Tr(\rho_\mathcal{A}^2)$ is the purity of the partition $\mathcal{A}$.
Note that the above potential is bounded by $1/d^{\floor{N/2}}\le \pi_{ME}\le 1$ and 
its lower bound is only saturated by AME states.

By minimizing the multipartite entanglement potential, explicit examples of AME states of 5 and 6 qubits are presented in Ref.~\cite{Facchi2008}. It is remarkable that even for a relatively
small number of qubits ($N\ge 7$), such minimization problem
has a landscape of
the parameter space with a large number
of local minima, what implies a very slow convergence.
The reason for that is frustration. The condition that purity saturates its minimum
can be satisfied for some but not for all the bipartitions (see \cite{FFMPP10b} for details).
In this respect, in \cite{FFMPP2010}, the minimization of the multipartite entanglement potential is mapped
into a classical statistical mechanics problem.
The multipartite entanglement potential is seen there as a Hamiltonian which is minimized by
simulated annealing techniques.

\section{Examples of AME states}\label{S3}

\subsection{ Qubit AMEs}

Let us consider states made out of qubits, that is the dimension of the local Hilbert space is $d=2$.
The simplest cases of AME states are any of the Bell states. There is a unique partition of two qubits and it is possible to entangle both parties maximally. 
It is easy to argue that 
there is a unique quantity that describes the amount of entanglement in the system. This can be chosen to be the first eigenvalue of the Schmidt decomposition of the state, or some other quantity derived from it as the von Neumann entropy. All two particle states, whatever their local dimension, can be entangled maximally. Those states are of no interest for our present discussion which is genuinely centered in multipartite entanglement.

In the case of 3 qubits the well-known GHZ state \cite{Dur00} is an AME, but it is known  that there is no 4-qubit AME \cite{HS00}. The amount of degrees of freedom in the definition of the state is insufficient to fulfill all the constraints coming from the requirement of maximum entanglement.

For 5 and 6 qubits, there are AME states. In particular, a 5-qubit state $\ket{\Upsilon_{5,2}} \in AME(5,2)$ can be defined by the coefficients of the superposition of basis states that form it:
\begin{equation}
\ket{\Upsilon_{5,2}} = \frac{1}{2^{5/2}}\sum_{i=0}^{2^5-1} c^{(\Upsilon)}_i |i\rangle,
\end{equation}
where we used the usual shorthand notation for the elements in the computational basis and the coefficients have the same modulus and signs given by \cite{Facchi2008}
\begin{eqnarray}
c^{(\Upsilon)}&=&\{1, 1, 1, 1, 1, -1,-1, 1, 1, -1,-1,\nonumber\\ 
&&1, 1, 1, 1,1,1, 1, -1, -1,1, -1,1,\nonumber\\ 
&&-1, -1, 1, -1, 1, -1, -1, 1, 1\} .
\label{ame52}
\end{eqnarray}
A state AME(5,2) found in \cite{Brown05} was proven to be useful for a number of multipartite tasks in Ref.
\cite{Muralidharan08}. It can also be found as the superposition
of a perfect error correcting code as presented in Ref. \cite{laflamme96} and also from orthogonal arrays \cite{GZ14}.

For the sake of completeness, let us also provide an absolutely maximally entangled 6-qubit state
\begin{equation}\label{eq:ame(6,2)}
\ket{\Xi_{6,2}} = \frac{1}{\sqrt 2^6}\sum_{i=0}^{2^6-1} c^{(\Xi)}_i |i\rangle,
\end{equation}
with 
\begin{eqnarray}
c^{(\Xi)}&=&\{
 -1,- 1,- 1,+ 1,- 1, 1, 1, 1, \nonumber\\ 
&&-1,- 1,- 1, 1,1,- 1,- 1,- 1, \nonumber\\ 
&&-1, -1, 1, -1,- 1, 1, -1,- 1, \nonumber\\ 
&&1, 1, -1, 1,- 1, 1, -1,- 1, \nonumber\\ 
&&-1, 1, -1,- 1,- 1,- 1, 1,- 1, \nonumber\\ 
&&1, -1, 1,1,- 1,- 1, 1,- 1, \nonumber\\ 
&&1, -1,- 1,-1, 1, 1, 1,- 1, \nonumber\\ 
&&1, -1,- 1,-1,- 1,- 1,- 1, 1 \} .
\label{ame62}
\end{eqnarray}

The case of 7 qubits remains unsolved
but numerical evidence hints at the impossibility of finding such an AME state
\cite{Borras07}. For 8 qubits or more, there are no AME states \cite{Sc04}.

\subsection{A central example: AME(4,3)}

The first non-trivial example of AME for larger local dimensions corresponds to
a state made of 4 qutrits, AME(4,3). Its explicit construction is
\begin{equation}\label{ame43}
\ket{\Omega_{4,3}}=
\frac{1}{3}\sum_{i,j,=0,1,2} |i\rangle|j\rangle |i+j\rangle |i+2 j\rangle
\ ,
\end{equation}
where all qutrit indices are computed mod(3). It is easy to verify that
all the reduced density matrices to two qutrits are equal to $\rho=\frac{1}{9} I_9$,
so that
this state carries entropy $S=2 \log 3$ for every one of its bi-partitions.

The state $\ket{\Omega_{4,3}}$ can be viewed as a map of a two-qutrit product basis into a second one. That is
\begin{equation}
  \ket{\Omega_{4,3}} = \sum_{i=0,8} |u_i\rangle |v_i\rangle
  = \sum_{i,j=0,8} |u_i\rangle U_{ij} |u_j\rangle
\end{equation}
where $\{|u_i\rangle\}$ and $\{|v_i\rangle\}$ are product basis for two qutrits, and  $|v_i\rangle = U_{ij} |u_j\rangle$.
The matrix $U_{ij}$ is not only unitary  (as it must as a consequence of multiunitarity) but
also a bijective map between the sets of words of length $2$ over an alphabet $\mathbb{Z}_3=\{0,1,2\}$.
In other words, the entries of $U_{ij}$ are $0$ with a single $1$ per row/column.
This property remains the same whatever partition is analysed, though the unitary will vary. 

A second feature of the state $\ket{\Omega_{4,3}}$ is that the Hamming distance between any pair of elements in the state is three, $D_H=3$. As all the sequences in $\ket{\Omega_{4,3}}$ differ in 3 elements, any single qutrit error can be corrected. Hence, maximal entanglement is related to error correction codes. Both of these properties are related to the fact that $|\Omega_{4,3}\rangle$ is an AME of minimal support, a concept we are going to explore in Sect. \ref{S4}.

It is also natural to expect AME states to accommodate easily to some magic-square like relations. A simple example goes as follows. Write the coefficients of $\ket{\Omega_{4,3}}$ as a matrix of row $i$ and column $i$, giving the composed value $a$ of the remaining two qutrits from 0 to 8, that is $a=3(i+j) {\rm mod}(3)+(i+2j) {\rm mod}(3)$. The square reads
\[
\begin{array}{ccc}
    $0$&  $5$& $7$\\
    $4$ & $6$&  $2$\\
    $8$ & $1$&  $3$
\end{array}
\]
where all rows and columns add up to 12. The same properties are maintain if we interchange the indices in the state. These kind of combinatorial designs are going to be explored in sect. \ref{S5}.

From an experimental point of view, $\ket{\Omega_{4,3}}$ can be created defining
a quantum circuit that generates it.
Such a circuit makes use of the following
two gates:
\begin{eqnarray}
  \label{eq:gates}
\nonumber
   {\rm Fourier} &\quad& F_3 |0\rangle=\frac{1}{\sqrt 3}\left(|0\rangle+
  |1\rangle+|2\rangle\right),
  \\
   {\rm C_3-adder}&\quad &U_{C-adder} |i\rangle|j\rangle=|i\rangle|(i+j) {\rm mod} 3\rangle.\nonumber\\
\end{eqnarray}
This gate $C_3$ generalizes the CNOT gate for qubits and it is represented in the circuit using the usual symbol for CNOT with the subscript 3.
\begin{figure}
\begin{center}
\scalebox{.3}{\includegraphics{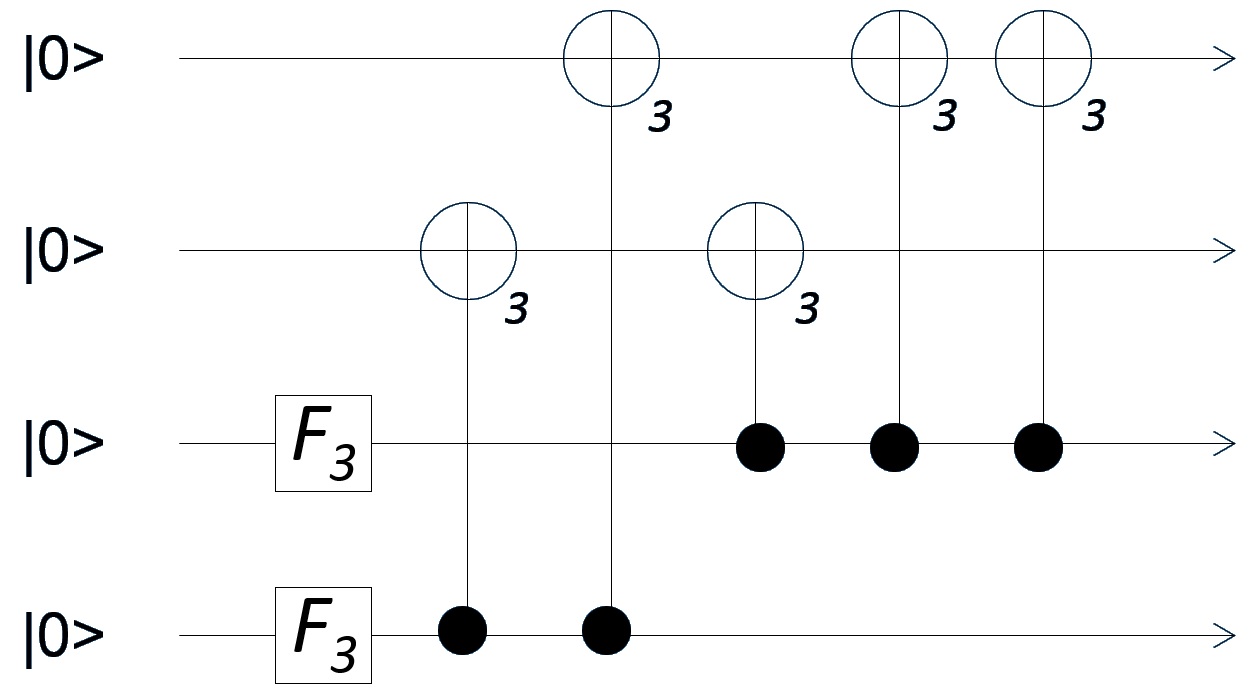}}
\end{center}
\caption{Quantum circuit required to generate the state AME $\Omega_{4,3}$ (4 qutrits) based on the Fourier gate $F_3$ and control-adders gate $C_3-adder$ .}
\label{fig:circuit2}
\end{figure}
The $\ket{\Omega_{4,3}}$ state  can be constructed as a
sequence of a Fourier $F_3$ and $C_3$-adders following the
circuit  depicted in Fig.\ref{fig:circuit2} acting
on the initial state $|0000\rangle$ of 4 qutrits.

\subsection{General expression for AME states}

The states AME(5,2) and AME(6,2) have the maximal number of terms whereas AME(4,3) has the minimal possible number. Therefore the following question arises: is there a general expression for AME($N,d$) having the maximal number of $d^N$ terms? Facchi partially solved this issue for qubits in \cite{Facchi2009}. It can be shown that a state expressed as
\begin{equation}
|\Psi\rangle= \sum_{k=0}^{2^{n}-1}  z_k |k\rangle , \hspace{5mm} z_k = r_k \xi_k ,
\end{equation}
where the $\ket{k}$ span the whole computational basis and $r_k$ and $\xi_k$ are respectively the modulus and the phase of the complex coefficients $z_k$, is an AME if it satisfies the following equations:

\begin{equation}
\label{eqphases}
\sum_{m}  r_{l \bigoplus m} r_{l' \bigoplus m} \xi_{l \bigoplus m} \xi_{l' \bigoplus m} = 0,
\end{equation}
where $l,l'$ and $m$ stand for both parts of a certain balanced bipartition (of $\floor{N/2}$ particles).

The general form of the squared modulus of the coefficients will be:

\begin{equation}
\label{eqmodulus}
r_k^2 = 2^{-N} + \sum_{\frac{N}{2} < n \leq N} \sum_{j \in S^n} c_j^{(n)} \prod_{1 \leq {h \leq n}} (2k_{h}^{(j,n)} -1),
\end{equation}
where $S^n$ stands for the set of bipartitions of the system into groups of $n$ and $N-n$ particles and $j$ is an index for the bipartitions of this set. $h$ is an index for particles contained in the bipartition $j$, and $k_{h}^{(j,n)}$ stands for the value (0 or 1) of particle $h$ in ket $k$, in the case of a certain bipartition $j$ of the set $S^n$. The real coefficients $c_j^{(n)}$ are free as long as they satisfy Eq.~(\ref{eqphases}) and the normalization condition.

With this expressions all AMEs of a small number of qubits are classified. It is also worth noting that AME states with maximal support and uniform amplitudes $r_k = 1 / \sqrt{d^N} \hspace{2mm} \forall k$ are possible to define (by just setting all $c_j$ to 0 in Eq.~(\ref{eqmodulus})) \cite{Facchi2009}.

\section{AME states of minimal support and classical codes}\label{S4}
In Ref.~\cite{Helwig-Cui2013}, a subclass of AME$(N,d)$ states 
is shown to be constructed by means of classical maximum-distance
separable (MDS) codes. In this section we show that such a subclass corresponds to the set of AME states of minimal support and exploit these ideas to get conditions for their existence.

\subsection{Support of AME states}

From the explicit examples we have presented in Sec. \ref{S3}, 
AME states appear to need a different numbers of elements to be written.
For instance, AME(4,3) is made of the superposition of 9 states, all
weighted with the same coefficient. Yet, AME(6,2) as written using
the 64 basis states with coefficients either 1 or -1.

Let us define the \emph{support} of a state $\ket{\psi}$ as 
the number of non-zero coefficients when $\ket{\psi}$ is written in the computational basis.
The \emph{support of a class} 
is defined as the support of the state inside the class with minimal support.
Note that the support of a class defines in turn another equivalence class.
Two states are \emph{support} equivalent if they belong to a LU classes
with equal support.

In this sense, it is interesting
to point out that the state AME(6,2), defined in Eq.~(\ref{eq:ame(6,2)})
with the maximal support of $2^6$ states, 
is LU equivalent to a state of support 16 by applying some Hadamard gates on its basis. 
It can be proven that 8 is the minimal support that such state could have but it is not attainable. 
Also in \cite{Brown05} a state AME(5,2) is built using 8 elements, while the theoretical minimum would be 4. 
We may wonder why the naive minimum possible number of $2^{\floor{N/2}}$ elements is not always attained. The answer to this question is answered in Ref.\cite{GZ14}, where it is proven a one-to-one relationship between $k$--uniform states having minimal support and a kind of combinatorial arrangements known as orthogonal arrays. Therefore, the non-existence of such states having minimal support is due to the non-existence of some classes of orthogonal arrays (those having strength one). In the following subsections we study in detail AME states having minimal support.

\subsection{Equivalence between AME states of minimal support and MDS codes}

A maximally entangled state in AME$(N,d)$ belongs to the class of minimal support iff
it is LU-equivalent to a state $\ket{\Psi}$ with support $d^{\floor{N/2}}$ i.~e.
\begin{equation}
\label{eq:tuples-state}
\ket{\Psi}= \frac{1}{d^{\floor{N/2}}}\sum_{k=1}^{d^{\floor{N/2}}} 
r_k\e^{\iu \theta_k}\ket{x_k}\, ,
\end{equation}
where $x_k\in \mathbb{Z}_d^N$ are words of length $N$
over the alphabet $\mathbb{Z}_d=\{0,\ldots, d-1\}$, and $r_k>0$ and $\theta_k\in [0,2\pi)$ are their modulus and phases respectively.

Given a bipartition $A=\{a_1,\ldots,a_{\floor{n}} \}$, it will be useful to introduce the subword $x_k[A]$ of the word $x_k$
for the partition $A$ as the concatenation of the $a_1$-th, $a_2$-th, $\ldots$ $a_n$-th letters of $x_k$, that is, 
$x_k[A]=x_k[a_1]x_k[a_2]\ldots x_k[a_n]$.
Let us also denote by $X_\Psi = \{x_k, k=1,\ldots,d^{\floor{N/2}}\}$ the set of words which $\ket{\Psi}$ has support on, and 
by $X_\Psi[A]=\{x_k[A], k=1,\ldots,d^{\floor{N/2}}\}$ the set of all subwords $x_k[A]$ corresponding to the bipartition $A$.
With this notation, the reduced density matrix of a partition $A$ can be written as
\begin{equation}
\rho_A=\sum_{k'}\bra{x_{k'}[\bar A]}\proj{\Psi} \ket{x_{k'}[\bar A]},
\end{equation}
where
\begin{equation}
\ket{\Psi}= \frac{1}{d^{\floor{N/2}}}\sum_{k=1}^{d^{\floor{N/2}}} r_k\e^{\iu \theta_k}\ket{x_k[A]}\ket{x_k[\bar A]}\, .
\end{equation}
In order for $\ket{\Psi}$ to be an AME, the reduced density matrix of 
any bipartition $A=\{a_1,\ldots,a_{\floor{N/2}} \}$ needs to be the completely mixed state.
It is easy to see that this has the following implications:
\begin{enumerate}
\item The modulus $r_k=1$ for all $k$.
\item The phases $\theta_k$ are arbitrary.
\item For any balanced bipartition $A$, with $|A|=\floor{N/2}$, 
two words $x_i,x_j\in X_\Psi$ have subwords $x_i[A]=x_j[A]$ if and only if $i=j$.
Equivalently, \ $X_\Psi[A]=\mathbb{Z}_d^{\floor{N/2}}$.
\end{enumerate}

Condition 3
implies that any pair of different words $x_i,x_j\in X_\Psi$ have \emph{Hamming
distance} 
\begin{equation}
d_H(x_i,x_j)\ge \floor{N/2}+1, 
\end{equation}
where the Hamming distance between two
codewords is defined as the number of positions in which they differ,
e.~g.\ $d_H(00010,10000)=2$.
To see this, note that otherwise there would exist
a balanced bipartition $A'$ for which $x_i[A']=x_j[A']$ for $i\neq j$ and
consequently the set $X_\Psi[A']$ would not contain all the possible words of length
$\floor{N/2}$, that is, $X_\Psi[A']\subset \mathbb{Z}_d^{\floor{N/2}}$.

A set of $M$ words of length $N$ over an alphabet of size $d$ that differ pairwise by at least a 
Hamming distance $\delta$ is called a \emph{classical code}. 
What we have shown above is that the existence of AME($N,d$) states of minimal support
imply the existence of classical codes of $M=|X_\Psi|=d^{\floor{N/2}}$ words with Hamming distance 
$\delta=\floor{N/2}+1$.
The codes produced by AME states are special in the sense that they saturate the \emph{Singleton bound} \cite{S64},
\begin{equation}
 M\le d^ {N-\delta+1}\, .
\end{equation}
This type of codes that saturate the Singleton bound are called maximum-distance separable (MDS) codes.

The converse statement is also true. That is, the existence of a MDS code also implies
the existence of an AME state with minimal support \cite{Helwig-Cui2013}.
The argument is the following: a code of $M=d^{\floor{N/2}}$ words of length $N$
 and Hamming distance $\delta=\floor{N/2}+1$
has all its subwords associated to any balanced bipartition $A$ of size $|A|=\floor{N/2}$ different, which implies the condition 3 above.
 Thus, \emph{AME states of minimal support are equivalent to
 classical MDS codes}.

This equivalence can be exploited to see that
a necessary condition for AME($N,d$) states with minimal support
(and equivalently of MDS codes) to exist is that the local dimension $d$ and the number of parties $N$ fulfill
\begin{equation}
d \ge \floor{N/2} +1 \, .
\label{dN21}
\end{equation}

We can prove it as follows: let us try to construct an AME state by building an MDS code.
Due to the relabeling freedom, the first $d+1$ words of the code can be chosen as
\begin{equation}
{\footnotesize
\begin{array}{c}
d+1\\
\textrm{code}\\
\textrm{words}
\end{array}}
\left\{
\begin{array}{rrr}
0&\ldots \ldots \ldots 00 & 0\ldots \ldots  \ldots 0 , \\ 
0&\ldots\ldots \ldots  01 & 1\ldots \ldots  \ldots 1 ,\\ 
\vdots & \vdots &   \vdots \\
0&\ldots 0 (d-1)& (d-1)\ldots (d-1),\\
0&\ldots \ldots\ldots 10 &  x_d[\floor{N/2}+1] \ldots x_d[N],
\end{array}
\right.\nonumber
\end{equation}
where the letters $x_d[i]$, for $\floor{N/2}+1 \le i \le N$, are still unknown
and every word is written in two subwords of lengths $\floor{N/2}$
and $\floor{(N+1)/2}$ respectively.

Note that:
({\it i}) none of the unknown letters $x_d[i]$ can be $0$ 
in order for the word $x_d$ to have Hamming distance $\floor{N/2} +1$ with the first word $0\ldots0 0\ldots0$.
({\it ii}) None of the unknown letters $x_d[i]$ can be repeated in order for $x_d$ to
have Hamming distance $\floor{N/2}+1$ with the other $d-1$ words.
Therefore, if $\floor{(N+1)/2}$ variables, $x_d[i]$,  must take $\floor{(N+1)/2}$ different values, 
and all of them must be different from 0,
it is necessary to extend the alphabet, forcing that $d\ge \floor{N/2}+1$.

Interestingly, Eq.(\ref{dN21}) forbids the existence of AME(N,2) states having minimal support for $N>3$. However, this does not represent a proof that AME(4,2) do not exist. Also, this inequality is saturated for the states AME(4,3) and AME(6,4). The existence of the cases AME(8,5) and AME(8,6) is still open whereas the states AME(8,7) and AME(8,8) are known.


\subsection{A less trivial example: AME(6,4)}
An application of the above connection between AME states and MDS codes is the construction
of AME states by exploring the set of all words and selecting those which differ in at least
$\floor{N/2}+1$ elements.
For the case of the AME(6,4), such a search gives a state with a equal superposition of the following entries:
\begin{eqnarray}
\label{AME4QQ}
&\{&000000, 001111, 002222, 003333, 010123, 011032, \nonumber \\
&&012301, 013210, 020231, 021320, 022013, 023102, \nonumber \\
&&030312, 031203, 032130, 033021, 100132, 101023, \nonumber \\
&&102310, 103201, 110011, 111100, 112233, 113322, \nonumber \\
&&120303, 121212, 122121, 123030, 130220, 131331, \nonumber \\
&&132002, 133113, 200213, 201302, 202031, 203120, \nonumber \\
&&210330, 211221, 212112, 213003, 220022, 221133, \nonumber \\
&&222200, 223311, 230101, 231010, 232323, 233232, \nonumber \\
&&300321, 301230, 302103, 303012, 310202, 311313, \nonumber \\
&&312020, 313131, 320110, 321001, 322332, 323223, \nonumber \\
&&330033, 331122, 332211, 333300\}\, .
\end{eqnarray}
%
%
Such an AME state carries the minimum possible support. 

\subsection{Construction of AMEs with minimal support}

Finding AME states by exploring the set of all words is highly inefficient
and becomes in practice unfeasible from a relatively small number of parties.
In this context, the Reed-Solomon codes \cite{RS60} can be a useful tool to produce
systematic construction of MDS codes and equivalently AME states.

Let us review here the particular case of $d$ prime and $N=d+1$.
Let us refer to the elements of the superposition
in the quantum states as words $x_i$ and the word of a half-partition
as $u_i$. The code words are obtained using
the action of a generator $G$, $x_i=u_i \cdot G$. The problem is then reduced to
fixing $G$. It can be shown that a family of valid generators is given by
\begin{equation}
  G=\left( 
  \begin{array}{ccccc}
  1&1&\ldots & 1& 0\\ 
  g_0 & g_1 &\ldots & g_{d-1}&0\\
  \ldots & \ldots & \ldots & \ldots & \ldots \\
  g_0^k & g_1^k &\ldots & g_{d-1}^k &1\\  
    \end{array}
  \right),
\end{equation}   
where $d\in \textrm{Prime}$, $N=d+1$, and $k=N/2$.

The case of AME(4,3) can be re-obtained using 
\begin{equation}
  G=\left( 
  \begin{array}{cccc}
  1&1 & 1& 0\\ 
  0 & 1 & 2 &1
   \end{array}
  \right)
\end{equation}
 Another concrete example corresponds to $g_0=1$, $g_1=1$,..., $g_6=6$ that
corresponds to AME(8,7), with $N=8$ $d=7$-dits. Then a total of $7^4$ codewords
are obtained that differ by a minimum Hamming distance $d_H=5$.

As the above construction can only be accomplished for $d$ being prime and $N=d+1$,
it is interesting to address the question of whether given some AME($N,d$) it is possible
to construct another AME($N',d'$). In this context, the following lemma can be useful which
allows to construct an AME($N',d$) for  any $N'<N$:
if there exists an AME state with minimal support in ${\mathcal H}(N,d)$ with $N$ being even, 
then there exist other AME states with minimal support in the Hilbert spaces 
${\mathcal H}(N', d)$ for any $N'\le N$.

To prove this last statement, 
we will consider separately the transitions $N\to N-1$ and $N-1\to N-2$. 

Transition $N\to N-1$:
The existence of an AME state with minimal support implies the existence of a
code of $d^{N/2}$ words of length $N$ with Hamming distance $d_H\ge N/2+1$. 
Let us order the words in the code in increasing order and take the subset of the first $d^{N/2-1}$ words which start with $0$. 
Note that by suppressing such a $0$, we get a code of $d^{N/2-1}$ words of length $N-1$ with Hammings distance
$N/2+1$, forming an AME.

Transition $N-1 \to N-2$:
From the previous steps, we are left with a code of $d^{N/2-1}$ words of length $N-1$ and
Hamming distance $N/2+1$. Note that by suppressing an arbitrary letter from all the words of the code,
one is left with a set of $d^{(N-2)/2}$ words of length $N-2$ with Hamming distance $N/2+1-1=(N-2)/2 +1$,
which is a MDS code. 

By iterating the previous procedure, we obtain MDS codes (and AME states of minimal support) for any $2\le N'\le N$. Equivalently, MDS codes are constructed from considering orthogonal arrays of strength one \cite{GZ14}.

\subsection{Non-minimal support AMEs and perfect quantum error correcting codes}

AME states are deeply related to classical error correction codes and compression \cite{laflamme96,Rains99,Sc04}.
This is somewhat intuitive since maximal entropy is related to maximally mixed subsets.
The measure of any local degree of freedom delivers an output which is completely random. This is, in turn, the basic element to correct errors. Hence, a relation between the elements superposed to form an AME state and error corrections codes is expected.

Le us illustrate the connection of an AME(5,2) state with the
well-known 5-qubit code \cite{laflamme96}. It is easy to see that by applying some Hadamard gates on local qubits, the AME(5,2) state, defined through
the 32 coefficient given in Eq. \ref{ame52}, is LU-equivalent to 
a state with fewer non-zero coefficients. Actually, a representative of
the same AME(5,2) class is found to have only 8 coefficients. That state corresponds to a superposition of the two
logical states in the error correcting codes found in \cite{laflamme96}
\begin{equation}
|\Omega_{5,2}\rangle = \frac{1}{\sqrt {2}} \left( |0\rangle_L + |1\rangle_L\right) ,
\end{equation}
where the logical qubits are defined as
\begin{eqnarray}
 |0\rangle_L &=& \frac{1}{2} \left( | 00000\rangle +|00011\rangle + |01100\rangle -|01111\rangle\right) \nonumber\\ 
 |1\rangle_L&=& \frac{1}{2} \left( | 11010\rangle +|11001\rangle+|10110\rangle -|10101\rangle \right)\nonumber
\end{eqnarray}
Note the fact that the coefficients carry both plus and minus signs
as is the case in the non-minimal support AMEs.

\section{Combinatorial designs and $k$-uniformity}\label{S5}

Combinatorial designs are arrangements of elements satisfying some
specific properties so called \emph{balanced} \cite{S04}. Such elements
are restricted to a finite set, typically considered as subsets of integer
numbers. Some remarkable examples are block designs, t-designs, orthogonal
Latin squares and orthogonal arrays (see \cite{Hedayat} and references
therein). Combinatorial designs have important applications in quantum
physics \cite{Z99}. Indeed, a connection between genuinely multipartite
maximally entangled states and orthogonal arrays has been recently found
\cite{GZ14}. Furthermore, they are a fundamental tool in optimal design of
experiments \cite{R88,ADT07}. The existence of some combinatorial designs
can be extremely difficult to prove. For example the existence of Hadamard matrices in every dimension multiple of four (i.e., the \emph{Hadamard
conjecture}) is open since 1893 \cite{H93} and it represents one of the most important open problems in Combinatorics.

\subsection{Relation to mutually orthogonal Latin squares}

Let us consider the explicit expression of the AME(4,3) state defined in  Eq.(\ref{ame43}):
\begin{eqnarray}
&\ket{\Omega_{4,3}} &=\frac{1}{3}\left( |0000\rangle+|0112\rangle+|0221\rangle\right. \nonumber\\
&&+|1011\rangle+|1120\rangle+|1202\rangle\nonumber\\
&&\left. +|2022\rangle+|2101\rangle+|2210\rangle\right).
\label{Pop33}
\end{eqnarray}
Here, the third and fourth symbols appearing in every term of the state can be arranged into 2 Greco-Latin squares of size three:
\begin{equation}
\label{greco}
\begin{array}{ccc}
A\alpha&B\gamma&C\beta\\
B\beta&C\alpha&A\gamma\\
C\gamma&A\beta&B\alpha
\end{array} 
\ = \ 
\begin{array}{ccc}
{ A\spadesuit}  & K\clubsuit& {Q\diamondsuit}\\
{ K\diamondsuit} & Q\spadesuit & { A\clubsuit}\\
{Q\clubsuit}    & A\diamondsuit  &{K\spadesuit}
\end{array},
\end{equation}
where every symbol of the sets $\{A,\alpha\}$, $\{B,\beta\}$ and $\{C,\gamma\}$, is associated to $0,1$ and $2$, respectively.
Hence a pair of symbols may represent a card of a given rank and suit.
Note that the first two digits in Eq.(\ref{Pop33}) may be interpreted as addresses
determining the position of a symbol in the square. Furthermore, by considering the following four Mutually Orthogonal Latin Squares (MOLS) \cite{CD96}
 of size five
\begin{equation}
\begin{array}{ccccc}
0000&4321&3142&2413&1234\\
1111&0432&4203&3024&2340\\
2222&1043&0314&4130&3401\\
3333&2104&1420&0241&4012\\
4444&3210&2031&1302&0123
\end{array},
\end{equation}
we define a 2-uniform state of 6 subsystems with five levels each:
\begin{widetext}
\begin{eqnarray}
|\Phi_5\rangle&=\frac{1}{5}&(|000000\rangle+|104321\rangle+|203142\rangle+|302413\rangle+|401234\rangle+\nonumber\\
&&|011111\rangle+|110432\rangle+|214203\rangle+|313024\rangle+|412340\rangle+\nonumber\\
&&|022222\rangle+|121043\rangle+|220314\rangle+|324130\rangle+|423401\rangle+\nonumber\\
&&|033333\rangle+|132104\rangle+|231420\rangle+|330241\rangle+|434012\rangle+\nonumber\\
&&|044444\rangle+|143210\rangle+|242031\rangle+|341302\rangle+|440123\rangle).
\end{eqnarray}
\end{widetext}
A state locally equivalent to $|\Phi_5\rangle$ has been previously found in \cite{GZ14} but its connection to MOLS is firstly given here.
 By considering the standard construction of maximal sets of $d-1$ MOLS of prime size $d$ we can generalize the above construction for quantum states of a prime number of levels $d$ and $N=d+1$ parties as follows
\begin{equation}
|\Phi_{d+1,d}\rangle=\frac{1}{d}\sum_{i,j=0}^{d-1}|i,j\rangle\bigotimes_{m=1}^{d-1}|i+jm\rangle.
\label{Popgen}
\end{equation}

It is well known that a maximal set of $d-1$ MOLS of size $d$ exist for every prime power $d=p^m$ \cite{Stin04}. This means that the above general expression can be extended to the case of prime power level systems. For instance, the maximal set of 3 MOLS of order $d=4$ can be represented by a color figure
\begin{equation}
\begin{array}{cccc}
{\color{blue}A\spadesuit}&{\color{green}K\diamondsuit}&{\color{orange}Q\heartsuit}&{\color{red}J\clubsuit}\\
{\color{orange}J\diamondsuit}&{\color{red}Q\spadesuit}&{\color{blue}K\clubsuit}&{\color{green}A\heartsuit}\\
{\color{green}Q\clubsuit}&{\color{blue}J\heartsuit}&{\color{red}A\diamondsuit}&{\color{orange}K\spadesuit}\\
{\color{red}K\heartsuit}&{\color{orange}A\clubsuit}&{\color{green}J\spadesuit}&{\color{blue}Q\diamondsuit}
\end{array}.
\end{equation}
This design determines an AME(5,4) given by
\begin{eqnarray}
|\Omega_{5,4}\rangle&=&|00000\rangle+|10312\rangle+|20231\rangle+|30123\rangle+\nonumber\\
&&|01111\rangle+|11203\rangle+|21320\rangle+|31032\rangle+\nonumber\\
&&|02222\rangle+|12130\rangle+|22013\rangle+|32301\rangle+\nonumber\\
&&|03333\rangle+|13021\rangle+|23102\rangle+|33210\rangle\nonumber,
\end{eqnarray}
where every symbol of the sets $\{A,\spadesuit,blue\}$, $\{J,\diamondsuit,orange\}$, $\{Q,\clubsuit,green\}$ and $\{K,\heartsuit,red\}$, is associated to $0,1,2$ and $3$, respectively,
while the first two digits of every term label the position of a symbol in the pattern. In the above expression a normalization factor is required. Note that this state, or a state equivalent with respect to local unitary transformations, 
arises from the Red-Solomon code of length five \cite{RS}. 

Furthermore, the construction can be extended to any dimension $d$ in the following way:
\begin{equation}
|\Phi_{d+1,d}\rangle=\frac{1}{d}\sum_{i,j=0}^{d-1}|i,j\rangle\bigotimes_{m=1}^{\mathcal{N}(d)}|\lambda_m[i,j]\rangle,
\label{ame43d}
\end{equation}
where $\mathcal{N}(d)$ denotes the maximal number of MOLS of size $d$.
Here $\lambda_m[ij]$ denotes the entries of the $m$-th Latin square,
so the above expression can be considered as a direct generalization of Eq.(2).
It is worth to add that for dimensions $d\ge 12$ not equal to a prime power number 
only lower bounds for the function  $\mathcal{N}(d)$ are known \cite{CD96}.
The problem is solved only for smaller dimensions, as
$\mathcal{N}(6)=1$ in agreement with unsolvability of the famous Euler problem of $36$
officers, while $\mathcal{N}(10)=2$ -- see \cite{Stin04} where explicit form
of a pair of MOLS of size ten is derived.
Thus for $d=10$ expression (\ref{ame43d}) describes a $2$-uniform state of
$4$ subsystems with $10$ levels each.

In general, the problem of constructing $N-2$ MOLS of size $d$ is equivalent to construct a 2-uniform state of $N$ qudits of $d$ levels having $d^2$ positive terms. Note that $d^2$ is the minimal number of terms that can have a 2-uniform state of qudits of $d$ level systems.

\subsection{AME states and hypercubes}
In the previous section we considered maximal sets of MOLS to construct 2-uniform states of qudits.
 However, this construction is not useful to find AME states for $d>4$. The aim of this section is to consider combinatorial arrangements for constructing AME states in such cases. The main result is inspired in a generalization of the AME(4,3) state $\ket{\Omega}$ given in Eq.(\ref{Pop33}) 
and the AME state of 6 ququarts presented in Eq.~(\ref{AME4QQ}).
\begin{figure}[!h]
\centering 
{\includegraphics[width=7.7cm]{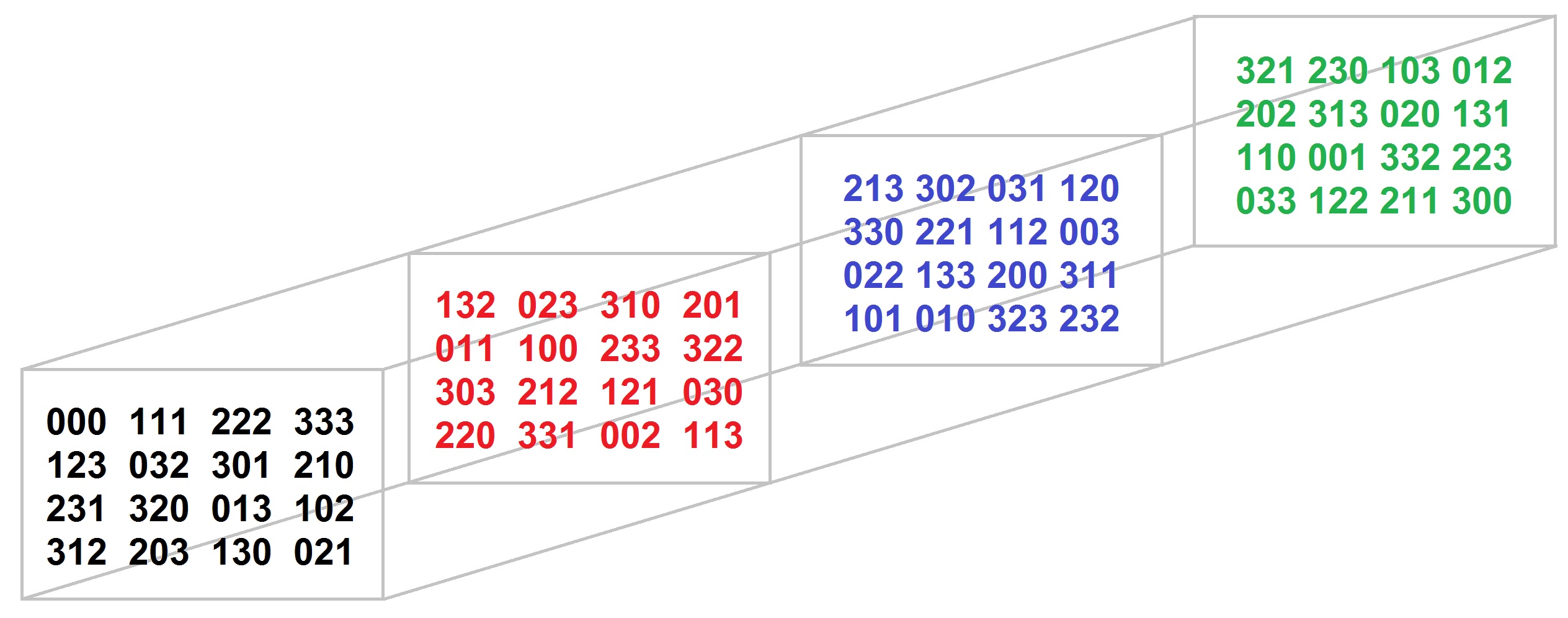}} 
\caption{Three mutually orthogonal Latin cubes of dimension $3$ and size $4$. 
This arrangement allows us to generate a state AME(6,4) of 6 ququarts.
 Each of the 12 planes (4 horizontal, 4 vertical and 4 oblique) contains a set of 3 MOLS of size 4.}
\label{Fig1}
\end{figure}
In \cite{GZ14} it was shown that this state can be derived from the irredundant orthogonal array IrOA$(64,6,4,3)$. Furthermore, this orthogonal array can be interpreted as a set of three mutually orthogonal Latin cubes (see Fig.\ref{Fig1}). Also note that the AME(4,3) state $\ket{\Omega}$ arises from an IrOA(9,4,3,2) (See Eq.(B1) in \cite{GZ14}). Thus, if $k$ mutually orthogonal hypercubes of dimension $k$ having $k+1$ symbols exist they are one to one connected with an IrOA($(k+1)^k,2k,k+1,k$) and, therefore, it would produce an AME(2k,k+1) state. This family of states would saturate the bound
\begin{equation}
d\geq\frac{N}{2}+1,
\end{equation}
already defined in Eq.(\ref{dN21}) Note that for $k=1$ one obtains the  standard Bell state, $|\Phi_1\rangle=(|01\rangle+|10\rangle)/\sqrt{2}$ (1 Latin square of size 2 with 2 symbols). Taking $k=2$ and using 2 MOLS of size 3 and 3 symbols we arrive to the AME(4,3) state of $\ket{\Omega}$ Eq.(\ref{ame43})  which, from this point of view, can be considered as a generalization of the Bell state. Furthermore, the state (\ref{AME4QQ}) corresponding to $k=3$ 
also belongs to this family and it is associated to three mutually orthogonal Latin 
cubes of dimension $3$ and size $4$. It is interesting to check,
whether there exist other states with $k\ge 5$ belonging to this family.

\section{AME states and multi-unitary matrices}\label{S6}

\subsection{Unitary matrices and bipartite systems}

Let us illustrate the connection between unitary matrices and AME states for the simplest case of  2 qubits. Let us assume that the state of the system is given by
\begin{equation}
|\phi\rangle=\frac{1}{\sqrt{2}}(U_{0,0}|00\rangle+U_{0,1}|01\rangle+U_{1,0}|10\rangle+U_{1,1}|11\rangle),
\end{equation}
where $\rho_A=\frac{1}{2}UU^{\dag}$, 
$\rho_B=\frac{1}{2}(U^T)(U^T)^{\dag}$  and $T$ denotes transposition. Any unitary matrix $U$ of size $2$ represents a Bell--like state.
Furthermore, the Pauli set of four unitary matrices $\mathcal{U}=\{\mathbb{I},\sigma_x,\sigma_y,\sigma_z\}$,
orthogonal in the sense of the Hilbert-Schmidt product, 
defines the maximally entangled Bell basis in ${\cal H}_2 \otimes {\cal H}_2$.

\subsection{Multi-unitary for AME(4,3)}
\label{sec:multi-unitarity}

We shall consider again the AME(4,3) state of four qutrits $\ket{\Omega_{4,3}}$, and 
represent its coefficients by a four--index tensor 
\begin{equation}
\ket{\Omega_{4,3}}=\sum_{i,j,k,l=0,1,2} t_{ijkl} \; |ijkl\rangle .
\label{tensorame43}
\end{equation}
where the entries of the tensor $t$ can be expressed as product of the Kronecker delta functions
\begin{equation}
 t_{ijkl}=\frac{1}{9}\delta_{k,i+j}\delta_{l,i+2j}.
\label{tensorcoef}
\end{equation}
Here, the addition operations are modulo 3. As discussed in the previous section, all non-zero coefficients are equal.
The tensor $t_{ijkl}$ consists of $3^4=81$ elements which can be reshaped
to form a square matrix of order $9$. Note that there exist altogether
$\binom{4}{2}=6$ different ways of choosing a bi-partition of the indices and forming a matrix $U_{\mu,\nu}$. That is,
\begin{equation}\label{munu}
(\mu,\nu) = \left\{
\begin{array}{c}
(i+2j,k+2l),\,(k+2l,i+2j)\\
(i+2k,j+2l),\,(j+2l,i+2k)\\
(i+2l,j+2k),\,(j+2k,i+2l)\\
\end{array}.
\right.
\end{equation}
The non-trivial property of an AME(4,3) tensor is that these six matrices are unitary. As transposition of a unitary matrix 
remains unitary, it is sufficient, in this case, to check
unitarity for the three cases appearing in the first column of the right side of Eq.(\ref{munu}). That is, taking combined indices in the original tensor,
\begin{equation}
  t_{ijkl}=U^{(1)}_{(ij)(kl)} = U^{(2)}_{(ik)(jl)} 
   = U^{(3)}_{(ij)(kl)} .
\label{multiunitarityame4}
\end{equation}
absolute maximal entanglement is achieved if the matrices  $U^{(1)}$, $U^{(2)}$ and $U^{(3)}$ correspond to different changes of bases, that is unitary matrices. 

We refer to such particular kind of unitary matrices as {\sl multi-unitary}. See Appendix \ref{appendixB} for further explanations about reorderings of elements in matrices of square size $D\geq4$.

\subsection{General multi-unitarity }
\label{sec:general-multi-unitaritity}

Let us consider a more 
general case of pure states of $N$ subsystems with $d$ levels each. That is,
\begin{equation}
|\phi\rangle=\sum_{s_0,\dots,s_{N-1}=0}^{d-1}
  t_{s_0,\dots,s_{N-1}}|s_0,\dots,s_{N-1}\rangle .
\end{equation}
Let us assume here that the number of subsystems is even, $N=2k$,
so there exist $M=\binom{2k}{k}$ possible splittings of the system
into two parts of the same size.

A necessary condition for $|\phi\rangle$ to be an AME state 
is that the tensor $t$ with $2k$ indices, reshaped into a square matrix of
size $d^{k}$ forms a unitary matrix $U$. 
This is so, as the reduction associated to the first $k$ qudits,
given by $\rho_{k}=UU^{\dag}$, should be proportional to the identity.
To arrive at an AME(2k,k) state, similar conditions have to hold
for all $M$ different square matrices obtained from the tensor $t$
by all possible ways of reshaping its entries into a square matrix. This observation provides a clear motivation to introduce 
the notion of {\sl multi-unitarity}:
A square matrix $A$ of order $d^k$ ($k\geq2$), acting on a composed Hilbert space ${\cal H}_d^{\otimes k}$ and represented in a product basis by 
$A_{\stackrel{\scriptstyle n_1, \dots , n_k}{\nu_1, \dots,  \nu_k}}:=
\langle n_1, \dots , n_k |A| \nu_1, \dots,  \nu_k\rangle$
is $k$--unitary if it is unitary for $M=\binom{2k}{k}$ reorderings of its entries corresponding to all possible choices of $k$ indices out of $2k$.

In this way, we can establish the following one-to-one connections:
\begin{equation*}
AME(2,d) \equiv \mbox{ unitary of order }d,
\end{equation*}
for bipartite systems of $2$ qudits having $d$ levels each and, in general:
\begin{equation*}\label{multiunitarityame43}
AME(2k,d) \equiv \mbox{ $k$-unitary of order }d^k,
\end{equation*}
for multipartite systems of $N=2k$ qudits having $d$ levels each. By construction, $1$--unitarity reduces to standard unitarity.
Any  $k$--unitary matrix with $k>1$ is called {\sl multi-unitary}. It is well-known that unitarity of matrices is invariant under multiplication. Multi-unitarity imposes more restrictions on a given matrix $U$ than unitarity. Therefore, the product of two
multi-unitary matrices in general is not multi-unitary. For instance, the matrix $O_8$ (see Eq.(\ref{O8}) below) is hermitian and 
$3$-unitary, but $O_8^2=\mathbb{I}$ is only $1$-unitary, as it represents a $6$--qubit quantum state equivalent to GHZ.

\smallskip

%
%
%
%
%


\medskip

Similarly, the case of the AME(4,3) state $\ket{\Omega}$ reduces to analyzing the properties of the tensor $t_{ijkl}$ in Eq. (\ref{tensorame43}) and verify the multi-unitarity of $U$. Indeed, for this state we have $U=Perm(0,5,7,4,6,2,8,1,3)$, $U^{T_2}=Perm(0,5,7,1,3,8,2,4,6)$ and $U^R=Perm(0,2,1,4,3,5,8,7,6)$, where $T_2$ and $R$ mean partial transposition and reshuffling (see appendix \ref{appendixB} for further details). Here, $Perm$ denotes a permutation matrix. A mini-catalogue of all the multi-unitary matrices defined in this work can be found in Appendix \ref{appendixD}.

\subsection{AME and Hadamard matrices}

For six qubits, the AME(6,2) state $\ket{\Xi_{6,2}}$ of Eq.(\ref{ame62}) having a maximal number of terms arises from graph states \cite{HEB04}. Let us write it explicitly
\begin{widetext}
\begin{eqnarray*}
\ket{\Xi_{6,2}}=\frac{1}{8}\left(\right. 
\!\!
&&-|000000\rangle -|000001\rangle -|000010\rangle +|000011\rangle -|000100\rangle +|000101\rangle+ |000110\rangle+ |000111\rangle\\ 
&&-|001000\rangle -|001001\rangle -|001010\rangle +|001011\rangle +|001100\rangle -|001101\rangle -|001110\rangle -|001111\rangle\\
&& -|010000\rangle -|010001\rangle + |010010\rangle -|010011\rangle -|010100\rangle +|010101\rangle -|010110\rangle -|010111\rangle\\
&& +|011000\rangle+ |011001\rangle -|011010\rangle+ |011011\rangle -|011100\rangle+ |011101\rangle -|011110\rangle -|011111\rangle\\
&& -|100000\rangle+ |100001\rangle -|100010\rangle -|100011\rangle -|100100\rangle -|100101\rangle +|100110\rangle -|100111\rangle\\
&&+ |101000\rangle -|101001\rangle +|101010\rangle +|101011\rangle -|101100\rangle -|101101\rangle +|101110\rangle -|101111\rangle\\ 
&&+|110000\rangle -|110001\rangle -|110010\rangle -|110011\rangle +|110100\rangle +|110101\rangle +|110110\rangle -|110111\rangle\\
&&\left. +|111000\rangle -|111001\rangle -|111010\rangle -|111011\rangle -|111100\rangle -|111101\rangle -|111110\rangle +|111111\rangle
\right).
\end{eqnarray*}
\end{widetext}
This state leads to the following  orthogonal matrix of order $D=2^3=8$
which is $3$--unitary:
\begin{equation}\label{O8}
O_8=\frac{1}{\sqrt{8}}\left(\begin{array}{rrrrrrrr}
-1& -1& -1& 1& -1& 1& 1& 1\\
-1& -1& -1& 1& 1& -1& -1& -1\\
-1& -1&  1& -1& -1& 1& -1& -1\\
1& 1& -1& 1& -1& 1& -1& -1\\
-1& 1& -1& -1& -1& -1& 1& -1\\ 
1& -1& 1& 1& -1& -1& 1& -1\\
1& -1& -1& -1& 1& 1& 1& -1\\
1& -1& -1& -1& -1& -1& -1& 1
\end{array}\right).
\end{equation}
Note that the entries of $\ket{\Xi_{6,2}}$ are given by the concatenation of the rows of $O_8$, up to normalization. This matrix is symmetric and equivalent up to enphasing and permutations \cite{TZ06}
to the symmetric Hadamard matrix $H_8=H_2^{\otimes3}$.

 Note that $O_8$ is 3-unitary but $H_2^{\otimes3}$ is not,
so permutation or enphasing  of a unitary matrix 
can spoil its multi-unitarity.
Moreover, from the concatenation of the rows of $H_2^{\otimes3}$ 
we only generate a 1-uniform state, which means that $H_2^{\otimes3}$ is only 1-unitary (i.e., unitary). 

We conjecture that
for any AME state one can choose suitable local unitary operations such that it is related to a multi-unitary complex Hadamard matrix. 
It represents a maximally entangled state with the maximal number of terms having all entries of the same amplitude. 
For example, the AME states arising from coding  theory \cite{RS60,Sc04} and graph states \cite{HEB04} are of this form. We recall that $k$-uniform states of $N$ qudits with $d$ levels having the minimum number of terms ($d^k$) are closely related to linear MDS codes (See Section 4.3 of \cite{Hedayat}) and also to orthogonal arrays of index unity \cite{GZ14}. This reasoning implies that
any pure state of $n$ subsystems with $d$ levels each having at least one reduction to $n/2$ qudits maximally mixed 
can have all its entries of the form $\pm d^{-n/2}$ if $d$ is even.

\subsection{Further constructions of AME states}\label{AMEconstruct}
Note that the state AME(4,3) $\ket{\Omega_{4,3}}$ is not equivalent with respect to local unitaries to a real state having all its entries of the form $\pm 3^{-2}$. This is a consequence of the fact that a real Hadamard matrix of size 9 does not exist. However, the state $\ket{\Omega_{4,3}}$ is equivalent under local unitary operations to the state (See Eq.(3) in \cite{Gaeta}):
\begin{equation}
\ket{\Omega'_{4,3}}=\frac{1}{9}\sum_{i,j,k,l=0}^2\omega^{j(i-k)+l(i+k)}|i,j,k,l\rangle,
\end{equation}
where $\omega=e^{2\pi i/3}$. This state is associated to the following $2$-unitary complex Hadamard matrix:
\begin{equation}
\label{Uame43}
U_{P}=\frac{1}{3}\left(\begin{array}{ccccccccc}
1&1&1&1&w&w^2&1&w^2&w\\
1&1&1&w^2&1&w&w&1&w^2\\
1&1&1&w&w^2&1&w^2&w&1\\
1&w&w^2&1&w^2&w&1&1&1\\
w&w^2&1&1&w^2&w&w^2&w^2&w^2\\
w^2&1&w&1&w^2&w&w&w&w\\
1&w^2&w&1&1&1&1&w&w^2\\
w^2&w&1&w&w&w&1&w&w^2\\
w&1&w^2&w^2&w^2&w^2&1&w&w^2
\end{array}\right).
\end{equation}
Interestingly, every integer power $(U_P)^m$ is a complex Hadamard matrix for $m\neq 4\,(\mathrm{Mod}\,4)$ and $(U_P)^8=\mathbb{I}$. 
Moreover, $U_P$ is equivalent to the tensor product of Fourier matrices $F_3\otimes F_3^{\dag}$, that is
\begin{equation}\label{equiv}
U_P=\mathcal{D}F_3\otimes F_3^{\dag}\mathcal{PD},
\end{equation}
where 
$\mathcal{D}=\mathrm{Diag}(1, 1, 1, 1,\omega,\omega^2, 1,\omega^2,\omega)$ is a diagonal unitary matrix,
while $\mathcal{P}$ is a permutation matrix which changes the order of the 
columns from $\{1,\dots,9\}$ to $\{1,4,7,2,5,8,3,6,9\}$.

In order to construct a $2$-unitary matrix one has to take a unitary $U$ such that its partially transpose $U^{T_2}$ and the 
reshuffled matrix $U^R$ are unitary -- see Appendix \ref{appendixB}. 
In the case of a matrix $U$ of size $D=3^2$ this implies that the set of nine $3\times3$ unitary matrices appearing in the $3\times3$ blocks of Eq.(\ref{Uame43}) define an orthogonal basis for the Hilbert-Schmidt product.
It is thus possible to obtain AME states by considering orthogonal bases of unitary operators.
For instance, one can construct  the $\ket{\Omega}$ state from the following matrix:
\begin{equation}\label{U_displa}
U^{\prime}_P=\frac{1}{\sqrt{3}}\left(\begin{array}{ccc|ccc|ccc}
1&0&0&1&0&0&1&0&0\\
0&1&0&0&\omega&0&0&\omega^2&0\\
0&0&1&0&0&\omega^2&0&0&\omega\\\hline
0&0&1&0&0&\omega&0&0&\omega^2\\
1&0&0&\omega^2&0&0&\omega&0&0\\
0&1&0&0&1&0&0&1&0\\\hline
0&1&0&0&\omega^2&0&0&\omega&0\\
0&0&1&0&0&1&0&0&1\\
1&0&0&\omega&0&0&\omega^2&0&0
\end{array}\right).
\end{equation}
We applied  here the orthogonal basis defined by the displacement operators of size $d=3$:
\begin{equation}\label{Displa}
D_{p_1,p_2}=\tau^{p_1p_2} X^{p_1}Z^{p_2},
\end{equation}
where $p=(p_1,p_2)\in\mathbb{Z}_d^2$, $\tau=-e^{\pi i/d}$, $\omega=e^{2\pi i/d}$, $X|k\rangle=|k+1\rangle$ and $Z|k \rangle=\omega^k|k\rangle$. These operators define the discrete Weyl-Heisenberg group. This approach can be easily generalized to any prime $d>2$. Indeed, for $d$ prime every reordering of indices lead us to the same matrix up to permutation of columns and rows, and therefore it remains unitary.
This shows a construction of AME($4$,$d$) working for any prime number of levels $d$.  Moreover, it is likely that this construction can be generalized for prime powers $d$ by considering the theory of Galois fields. 
Observe that the above construction is essentially different from the construction of AME states used in coding theory. 
Indeed, the tensor products of $N$ displacement operators bases of size $d$, i.e. the set $\{D_{p^1_1,p^1_2}\otimes \dots\otimes D_{p^N_1,p^N_2}\}$, produce codes and states AME($N$,$d$) \cite{Sc04}.

\section{Conclusion}\label{S7}

In this work we have analyzed some new properties of Absolutely Maximally Entangled (AME)
states in multipartite systems. First of all, we have reviewed and extended
several ways of constructing them. Then, we have explored their relation to the
field of combinatorial designs. For instance, 
a state AME(4,3) consisting of four maximally entangled qutrits
is linked to the set of two mutually orthogonal
Latin squares of order $3$, while a state AME(6,4) made of 6 ququarts is related to the set
of three mutually orthogonal Latin cubes of order $4$.

A profound relation between AME states and tensors that display the property of multi-unitarity has been found: AME states made out of an even number $N$ of degrees of freedom are equivalent to multi-unitary matrices (i.e., matrices being unitary after $M=\binom{N}{N/2}$ rearrangements of its entries). This remarkable property may be at the core of the use of AME states in holography \cite{LS15,PYHP15}. 

Furthermore, making use of a state AME($2k$,$d$) consisting of $N=2k$ parties, one can construct a $k-1$-uniform state state of $2k-1$ parties by removing a single subsystem.

We have proven the existence of states AME(4,$d$) for every prime $d>2$. Note that this also provides the existence of states AME(3,$d$) for every prime $d>2$.

AME states remain mainly unexplored. Let us bring a number of open problems
that deserve to be solved.
\begin{itemize}
\item 7 qubits.\\
Is there an AME state of 7 qubits?

\item Classification of AME and LU invariants.\\
It is unclear whether there is a clear cut classification of AME, which is related to
LU invariants. The fact that some AME states carry different minimal supports or that
some AME are right away related to Reed-Solomon codes hints a some unknown structure
among AME states.

\item Non minimal support.\\
Examples of AME states with non-minimal support are 
only explicitly known for 5 and 6 qubits. It would
be natural to find examples for higher local dimensions.

\item Computation of invariants.\\
LU invariants grow exponentially with the size of the number of parties.
An example of them is the hyperdeterminant, which has only been computed up to
4 qubits. There are no computations of hyperdeterminants of 4 qutrits.
Do AME states carry maximum values for some LU invariant? From the
general theory of hyperdeterminants, we know that the rank of the tensor
defining a four-qutrit state has rank 1269, which seems out of reach for 
any practical computation.

\item Bell inequalities.

Bell inequalities related to AME(4,3) are unknown. 

\end{itemize}

The small corner of the Hilbert space formed by AME states may
well be more complex than expected.

\bigskip

{\bf Acknowledgements}.
It is a pleasure to thank J.~Bielawski,  M.~Grassl and S.~Pascazio
for useful remarks. DG is thankful to Pawe{\l} Horodecki for the hospitality during his stay in Sopot, where
this project was initiated. JIL acknowledges financial support by FIS2013-41757-P and DA the APIF grant from Universitat de Barcelona. AR thanks support from the Beatriu de Pin\'os fellowship (BP-DGR 2013) and EU IP SIQS. This work was supported by the ERC Advanced Grant  QOLAPS coordinated by Ryszard Horodecki. 

\appendix

\section{Four-qubit entangled states}
\label{appendixA}


For completeness we will present in this appendix a discussion 
of maximally entangled states of four qubits. 
To see that no AME(4,2) states exist it is sufficient to analyze
the purity of reduced density matrices.
The purity ${\rm Tr} \rho^2$, serves as a measure of the degree of mixedness
of the density matrix $\rho$,
but also as a measure of entanglement of the initially pure state reduced to $\rho$
by a partial trace. 

The same argument works for multipartite systems.
 Let us denote the qubits by A, B, C and D. 
The purity of the bipartition AB of a state $\ket{\psi}$ is given by  $\Tr( \rho^2_{AB})$ with
$\rho_{AB}=\Tr_{CD}{\ketbra{\psi}{\psi}}$ being the reduced density matrix of AB. 
The theoretical minimum for this quantity is $1/N$; so $1/4$ in the case of four qubits. However, one cannot attain this minimum for the three bipartitions at the same time, as it was proven analytically in Ref.~\cite{HS00}. 
The best one can have is 1/3 in all bipartitions. 
One state that accomplishes this was given in that same paper:
\begin{eqnarray}
  |HS\rangle&=&\frac{1}{\sqrt{6}}\left[|0011\rangle + |1100\rangle +\omega |0101\rangle + \nonumber\right.\\
  &&\left.\omega |1010\rangle + \omega^2 \left(|0110\rangle + |1001\rangle \right)\right],\nonumber\\
 \end{eqnarray}
 where $w=\rm{exp}(\frac{2i\pi}{3})$. This state has also maximum entropy of entanglement.
Another state that has minimum purity in all bipartitions is:
\begin{eqnarray}
  | HD \rangle &=& \frac{1}{\sqrt 6} (|0001\rangle + |0010\rangle + |0100\rangle \nonumber\\
  && +|1000\rangle +\sqrt{2} |1111\rangle).
 \end{eqnarray}
This state was found by the authors and by \cite{Osterloh05} to also carry maximum hyperdeterminant, an extension of the concept of determinant to higher dimensions \cite{Gelfand94}, and an interesting entanglement measure that is intrinsically multipartite. An equivalent state already appeared in \cite{GBP98} as an example of a "symmetric maximally entangled state" of 4 qubits.

Let us note that AME states are defined through their entanglement properties and thus can be transformed into any equivalent state under local unitaries. Gour and Wallach \cite{Gour10} found $|L\rangle$ and $|M\rangle$ states while searching, respectively, for the states 
that maximize the average Tsallis $\alpha$-entropy of entanglement for $\alpha >2$ and for $0<\alpha <2$:
\begin{widetext}
\begin{align}
   |L\rangle =&\frac{1}{\sqrt{12}}\left(\left((1+\omega)(|0000\rangle + |1111\rangle\right)+(1-\omega)\left(|0011\rangle + 
   |1100\rangle\right)\right. \nonumber \\
   &\left.+\omega^2\left(|0101\rangle + |0110\rangle + |1001\rangle + |1010\rangle\right)\right) \\
   |M\rangle=&\frac{1}{\sqrt{2}}\left(\left(\frac{i}{2}+\frac{1}{\sqrt{12}}\right)\left(|0000\rangle + |1111\rangle\right) + \left(\frac{i}{2}-\frac{1}{\sqrt{12}}\right)\left(|0011\rangle + |1100\rangle\right)\right. + \nonumber \\
&\left.\frac{1}{\sqrt{3}}\left(|0101\rangle + |1010\rangle\right)\right),
\end{align} 
\end{widetext}
where $\omega=e^{2\pi i/3}$.
Remarkably, states $|L\rangle$ and $|M\rangle$ can be transformed by SLOCC into the states $| HD \rangle$ and $|HS\rangle$ respectively. They are called SLOCC equivalent.

In fact Verstraete et al. gave a classification of all pure 4-qubit states in 9 SLOCC inequivalent classes \cite{Verstraete02}. The most important class is called the generic class and is presented by Gour and Wallach \cite{Gour10} in the following compact form:
\begin{eqnarray}
\mathcal{G} &\equiv& \lbrace z_0|\phi^+\rangle|\phi^+\rangle
+z_1|\phi^-\rangle|\phi^-\rangle
+z_2|\psi^+\rangle|\psi^+\rangle\nonumber\\
&&+z_3|\psi^-\rangle|\psi^-\rangle \mid z_0,z_1,z_2,z_3 \in \mathbb{C} \rbrace ,
\end{eqnarray}
where $|\phi^\pm\rangle=(|00\rangle \pm |11\rangle / \sqrt{2}$ and $|\psi^\pm\rangle=(|01\rangle \pm |10\rangle / \sqrt{2}$ 
are the Bell states.

The method to transform a state from the computational basis into the Bell basis (which means obtaining the values of the 
$z_i$) is explained in detail in Verstraete's paper. Any 2 states that have the same 4 parameters $z_0, z_1, z_2$ and $z_3$ in this basis are SLOCC-equivalent. 
The SLOCC-equivalence between the states $|L\rangle$ and $| HD \rangle$ (as well as between the states $|M\rangle$ and $|HS\rangle$) is proven by showing that they have
precisely the same $z_i$ coefficients.

\section{Partial transposition, reshuffling and 2-unitarity}
\label{appendixB}

For any matrix $X$ of a square order $D=d^2$
represented in  a product basis, 
$X_{\stackrel{\scriptstyle mn}{\mu,\nu}}:=
    \langle m,n|X| \mu,  \nu\rangle$,
one defines  \cite{ZB04}
its partial transposition,
$X_{\stackrel{ \scriptstyle m n }{\mu \nu}}^{T_2}
=X_{\stackrel{ \scriptstyle m \nu }{\mu n}}$,
and reshuffling,
$X_{\stackrel{ \scriptstyle m n }{\mu \nu}}^{R}
=X_{\stackrel{ \scriptstyle m \mu }{n \nu}}$.
To get a better feeling of these particular reorderings
of elements we consider a matrix $X$ of order four.
Let us now switch to the standard, two--index notation,
and write its elements as $X_{ij}$ with $i,j=1,2,3,4$.
Here, we have $\binom{4}{2}=6$ different reorderings of 2 indices out of 4. Two of these reorderings are particularly interesting: \emph{(a)} the partially transposed matrix $X^{T_2}$
is equivalent to the matrix with all four blocks of size two transposed, and \emph{(b)}
the reshuffled matrix $X^R$ is obtained by
taking lexicographically each  $2 \times 2$ block of $X$,
reshaping it into a vector of length $4$ and putting into 
the reordered matrix $X^R$. That is,   
\begin{eqnarray}
X^{T_2} :=\left[
\begin{array}{c|c}
 X_{11} {\bf {\color{blue} \ \ X_{21}}}    & X_{13}{\rm ~ ~ ~ }{\bf {\color{red} X_{23}}} \\
 {\bf {\color{blue} X_{12}}} {\rm ~ ~ ~ }X_{22} & {\bf {\color{red} X_{14}}} \ \ X_{24} \\
\hline
 X_{31} {\ \ \bf {\color{green} X_{42}}} & X_{33} {\rm ~ ~ ~ }{\bf  {\color{orange} X_{43}}} \\
{\bf {\color{green} X_{24}}}{\rm ~ ~ ~ }X_{42} & {\bf  {\color{orange}  X_{34}}}\ \ X_{44} 
\end{array}
\right],\nonumber\\
X^R :=\left[
\begin{array}{c|c}
{ {X_{11}\ \ X_{12}}}  & {\bf {\color{blue} X_{21}} {\rm ~ ~ ~ } {\color{red} X_{22}}} \\
{\bf {\color{blue} X_{13}} {\rm ~ ~ ~ } {\color{red} X_{14}}} & { X_{23} \ \ X_{24}} \\
\hline
{ X_{31}\ \ X_{32}} & {\bf {\color{green} X_{41}} {\rm ~ ~ ~ }  {\color{orange}  X_{42}}} \\
{\bf {\color{green} X_{33}}{\rm ~ ~ ~ } {\color{orange}  X_{34}}} & { X_{43}\ \ X_{44} }
\end{array}
\right] .
\label{reshuf1}
\end{eqnarray}
Here, colored exchanged entries are set in boldface (colors in online version). The other 3 matrices are transpositions of $U$, $U^{T_2}$ and $U^R$. Therefore, a matrix $U$ of size 4 is multi-unitary if $U$, $U^{T_2}$ and $U^R$ are unitary. The same restrictions hold for matrices of size $d^2$.

It is possible to demonstrate that 2-unitary matrices of size $D=4$ do not exist \cite{HS00}. 
The smallest 2-unitary matrix exists for order $D=9$ -- see Eq.(\ref{Uame43}).


\section{Symmetric sudoku, generalized AME(4,3) states and $2$--unitary permutations}
\label{appendixC}
According to the usual sudoku rules all digits in a single row/column of the matrix
 or block of size three are different.  
Let us distinguish other sets of nine elements:
{\sl location}, which contain all digits from the same place 
in each block (e.g. nice centers of the blocks - set in red in the matrix below);
 {\sl broken rows} containing three rows of length three occurring in the same 
position of three blocks (blue example in the matrix) 
and an analogous notion of {\sl broken column}.

Note that the standard operation of matrix transpose,  $X^T$, exchanges columns of the
matrix with its rows. The partial transposition $X^{T_2}$ exchanges rows with broken columns
and columns with broken rows. Furthermore, reshuffling operation, $X^R$
interchanges blocks with rows and columns with locations -- see (\ref{reshuf1}),


The following matrix shows an example of 
a {\sl symmetric sudoku} pattern analyzed in \cite{BCC}:
all digits in each column, each row, each block, each {\color{red} location (red in online version)},
each {\color{blue} broken row (in blue)}, and 
each  {\bf broken column (boldface)} are different.

{\renewcommand{\arraystretch}{0.65}
\begin{equation}
\label{sudok9}
S_9 :=\left[
\begin{array}{
  c c c     |  c c c       | c c c }
\hline
 {\color{blue} \bf 8} & {\color{blue} 1} & {\color{blue}6}  &  {\bf 2} & 4 & 9   & {\bf 5} & 7 & 3 \\
 {\bf 3} & {\color{red} 5} & 7     &  {\bf 6} & {\color{red}8} & 1   & {\bf 9} & {\color{red} 2} & 4 \\
 {\bf 4} & 9 & 2  &  {\bf 7} & 3 & 5   & {\bf 1} & 6 & 8 \\
\hline
{\color{blue} 7} & {\color{blue}3} & {\color{blue} 5}  &  1 & 6 & 8   & 4 & 9 & 2 \\
 2 & {\color{red}4} & 9  &  5 & {\color{red}7} & 3   & 8 & {\color{red}1} & 6 \\
 6 & 8 & 1  &  9 & 2 & 4   & 3 & 5 & 7 \\
\hline
{\color{blue} 9} & {\color{blue} 2} & {\color{blue} 4}  &  3 & 5 & 7   & 6 & 8 & 1 \\
 1 & {\color{red} 6} & 8  &  4 & {\color{red}9} & 2   & 7 & {\color{red}3} & 5 \\
 5 & 7 & 3  &  8 & 1 & 6   & 2 & 4 & 9 \\
\hline
\end{array}
\right] .
\end{equation}
}

\medskip
Consider now a matrix $\mathcal{P}_1$ of size nine with all entries equal to zero
besides nine elements placed in the positions of digit '1' of the matrix
(\ref{sudok9}) equal to unity. As all nine digits in each location of $S_9$
are different it is clear that this is a legitimate permutation matrix
of size $9$. As the address of each non-zero element can be 
interpreted as a pair of two ternary digits, it represents
a Graeco-Latin square (\ref{greco})
and determines the AME(4,3) state $\ket{\Omega}$ (\ref{Pop33}).

Thus permutation matrix $\mathcal{P}_1$ is $2$--unitary, as its partial transpose and reshuffling 
remain unitary. The same property holds also for other permutation matrices $\mathcal{P}_m$ 
with $m=2,\dots, 9$, obtained by placing nine ones in the positions occupied by 
digits 'm' in  pattern (\ref{sudok9}). The property of $2$-unitarity is preserved if we
enphase a permutation matrix by multiplying it by a diagonal unitary matrix $\mathcal{D}$,
and take $\mathcal{P}_m'=\mathcal{DP}_m$. Thus it is fair to say that the symmetric sudoku 
matrix (\ref{sudok9}) encodes nine families of enphased $2$--unitary permutation matrices
or, families of AME(4,3) states.

\section{Mini catalogue of multi--unitary matrices}
\label{appendixD}

Multi--unitary matrices are defined for orders equal to powers of integers. 
As there are no $2$-unitary matrices of size $4$ the smallest interesting cases 
are $D=8,9$ and $D=16$. 
In general, multi-unitary permutation matrices are one-to-one connected \cite{GZ14} to combinatorial arrangements 
called orthogonal arrays of index unity \cite{CD96}.
Here, we present a resume of all the multi-unitary matrices presented in this work and some additional ones:
\medskip 

\begin{itemize}
\item[(\emph{i})] $D=2^3=8$. The orthogonal matrix $O_8$ defined in (\ref{O8})
is $3$--unitary. This matrix is equivalent with respect to permutations and 
enphasing to the three--qubit Hadamard matrix $H_8=H_2^{\otimes 3}$ (which is 1--unitary). 
It is not possible to construct a $3$--unitary permutation matrix of size $D=8$ \cite{GZ14}.
 
\item[(\emph{ii})] $D=3^2=9$. There exist $2$--unitary permutation matrices related to symmetric sudoku designs  -- see Appendix \ref{appendixC}. Also permutation matrices defined at the end of Section \ref{sec:multi-unitarity} and the complex Hadamard 
matrix (\ref{Uame43}) are $2$-unitary. The latter one is equivalent to the tensor product of two Fourier matrices $F_3 \otimes F_3$,
but this product is 1--unitary only. 

\item[(\emph{iii})] $D=2^4=4^2=16$. In this case there are no $4$--unitary matrices  \cite{Sc04}. 
There exist a $2$--unitary permutation matrix given by {\small$Perm(4,3,13,10,14,9,7,0,11,12,2,5,1,6,8,15)$} created from OA(16,4,4,2), 
which was generated using the Gendex Module NOA \cite{NOA}.

\item[(\emph{iv})] $D=p^2$ for a prime $p$. Two-unitary matrices of this size exist, but they are neither complex Hadamard matrices ($d^4$ non-zero entries) nor permutations ($d$ non-zero entries). Indeed they have $p^2$ non-zero entries which correspond to powers of the main root of the unity $e^{2\pi i/d}$. \emph{Construction:} Let us denote $(j,k)$ to the $j$-th row block of size $p$ and the $k$-th column block of size $p$ of $U$ (note that $U$ has $d^2$ blocks of size $p$). Each block $(j,k)$ has to be filled with the displacement operator $D_{j,k}$ of size $d$. Displacement operators are defined in Eq.(\ref{Displa}). The explicit $U$ for $d=3$ is provided in Eq.(\ref{U_displa}). 
\end{itemize}

As we can see, for some dimensions there are no 2--unitary permutation matrices. Indeed, a 2--unitary matrix of size $d^{(d+1)/2}$ exists if and only if a projective plane of odd order $d$ exist (see Theorem 8.43 in \cite{Hedayat}). In particular, they exist for every prime (and odd prime power) $d$. For example, the existence of AME(4,3) states relies on the existence of the projective plane of order 3.

\end{document}